\documentstyle{article}
\begin{document}

\title{Quantum key distribution with authentication}
\author{Guihua Zeng \thanks{Email: ghzeng@pub.xaonline.com},\,\,\,\,\,\,\,\, Xinmei Wang\\
National Key Lab. on ISDN, XiDian University, Xi'an 710071, P.R.China\\}
\date{}
\maketitle
\begin{abstract}
The security of the previous quantum key distribution (QKD) protocols, which is
guaranteed by the nature of physics law, is based on the legitimate users.
However, impersonation of the legitimate communicators by eavesdroppers, in practice, will
be inevitable. In fact, the previous QKD protocols is un secure without authentication in 
practical communication. In this paper, we proposed an improved QKD protocol
that can  simultaneously distribute the quantum secret key and verify the communicators'
identity. This presented authentication scheme is provably secure.\\
PACS:0365.Bz\\[1.0cm]
\end{abstract}
\large

\section{Introduction}

Since the first finding that quantum effects may protect privacy information transmitted 
in an open quantum channel by S.Wiesner [1], and then  by C.H.Bennett and G.Brassard [2],
a remarkable surge of interest in the international scientific and industrial
community has propelled quantum cryptography into mainstream computer science
and physics. Furthermore, quantum cryptography is becoming increasingly
practical at a fast pace. Several quantum key distribution protocols have been 
proposed, three main protocols of these are the BB84 protocol [3], B92 protocol [4], and 
EPR protocol [5]. The first quantum key distribution prototype, working over a distance 
of 32 centimetres in 1989, was implemented by means of laser transmitting in free space [6].
Soon, several experimental demonstrations by optical fibre were set up[7]. After that, 
a lot of publications have been presented, which cover three 
aspects: 1) QKD protocols [8-10], 
2) security of QKD protocols and detection of eavesdropper [11-16] 
and 3) practical application of quantum cryptography [17-20].

Quantum cryptography  employs
quantum phenomena such as the Heisenberg uncertainty principle and the quantum
corrections to protect distributions of cryptographic keys. Key distribution
is defined as procedure allowing two legitimate users of communication channel
to establish two exact copies, one copy for each user, of a random and secret
sequence of bits. In other words,
quantum cryptography is a technique that permits two
parties, who share no secret information initially, to communicate over an
open channel and to establish between themselves a shared secret sequence of
bits. Quantum cryptography is provably secure against eavesdropping attack,
in that, as a matter of fundamental principle, the secret data can not be
compromised unknowingly to the legitimate users of the channel.
Three ingenious protocols in quantum cryptography have been proposed. The
first, by Bennett {\it et al} [3], relies on the uncertainty principle of quantum
mechanics to provide key security. The security guarantee is derived from
the fact that each bit of data is encoded at random on either one of a
conjugate pair of observables of quantum-mechanical object. Because such a
pair of observables is subjected to the Heisenberg uncertainty principle,
measuring one of the observables necessarily randomizes the other. A further
elegant technique has been proposed by Ekert [5], which relies on the violation
of the Bell inequalities to provide the secret security. And the third
technique, devised by Bennett [4], is based on the transmission of nonorthogonal
quantum states.

However, the cryptography and the cryptanalytics are always a pairs contradiction. Once a 
cryptographic protocol is proposed, eavesdropper (Eve) will try to break it. 
Quantum cryptography is also no exception, although the quantum cryptography is thought to 
be provably security. With the quantum key distribution protocols presented, 
several attacks strategy have been proposed, such as intercept/resend scheme [6], 
beamsplitting scheme [6], entanglement scheme [14] and quantum copying [15,16]. 
In the intercept/resend scheme, Eve intercepts selected light pulses and reads them 
in bases of her choosing. When this occurs, Eve fabricates and sends to Bob a pulse of 
the same polarization as she detected. However, due to uncertainty principle, at least 
25\% of the pulse Eve fabricates will yield the wrong result if later successfully measured 
by Bob. The other attack, beamsplitting, depends on the fact that transmitted light pulses 
are not pure single-photon states. In the entanglement scheme, Eve involves 
the carrier particle in an interaction with her own quantum system, referred to as probe, 
so that the particle and the probe are left in an entangled state, and a subsequent 
measurement of the probe yields information about the particle. Some investigators are 
now turning their attention to collective attacks and joint attacks. About these 
attacks description please see Ref.[13] and its references. Eve can also use the 
quantum copying to obtain information between Alice and Bob. Two kind quantum copies are 
presented [15,16]. It is appropriate to emphasize the limitation of above attacks strategy.
All these mentioned attacks strategy are restricted by the uncertainty principle or the 
quantum corrections, so Eve can not break the quantum cryptography protocols.
The risk of eavesdropper is to disturb the information and to be detected finally  by 
the legitimate users. This is the reason why quantum cryptography is declared to be 
provably secure.

However, The security of the previous quantum key distribution protocols, which is
guaranteed by the nature of physics law, is based on the legitimate users.
In practice, the impersonation of Alice or Bob by eavesdropper will
be existed in a large probability. Circumventing the impersonation needs to
verify the identity of communicators, saying Alice and Bob. 

In section 2 we give the preliminary ingredients, which are important quantum effects in 
quantum cryptography. Section 3 reviews the QKD protocols. In section 4 we propose a QKD 
protocol with authentication function. It is composed of two phases, i.e., the initial phase
and QKD with authentication phase. In section 5 we analyze the security of the presented 
scheme. We conclude in section 6.

\section{Preliminaries}

The recent results in quantum cryptography are associated with the Heisenberg
uncertainty principle of quantum mechanics, EPR effects. In follows, we briefly 
describe these quantum effects.

\subsection{Heisenberg Uncertainty Principle}

Using standard Dirac notation, this principle can be succinctly stated as follows:
\begin{description}
\item Heisenberg Uncertainty Principle: For any two quantum
mechanical observables $A$ and $B$, if the corresponding operators $\hat {A}, \hat {B}$ satisfy
$$\left[\hat {A},\hat {B}\right]  =\hat{A}\hat{B}-\hat{B}\hat{A},$$
then
\[
\left\langle \Delta A\right\rangle \left\langle \Delta B\right\rangle 
\geq\frac{1}{2}\left\|  \left\langle
\left[  A,B\right]  \right\rangle \right\| {,}
\]
where $\left\langle \Delta A\right\rangle$ and $\left\langle \Delta B\right\rangle$ are 
average value of operators $\Delta \hat{A}$ and $\Delta \hat{B}$, respectively.
\[
\Delta \hat{A}=\hat{A}-\left\langle A\right\rangle {\qquad and\qquad}\Delta
\hat{B}=\hat{B}-\left\langle B\right\rangle {.}
\]
\end{description}

Thus, $\left\langle \Delta A\right\rangle $ and
$\left\langle \Delta B \right\rangle $ are variances which
measure the uncertainty of observables $A$ and $B$. For incompatible
observables, i.e., for observables $A$ and $B$ such that $\left[  A,B\right]
\neq0$, reducing the uncertainty $\left\langle \Delta A
\right\rangle $ of $A$ forces the uncertainty $\left\langle \Delta
B\right\rangle $ of $B$ to increase, and vice versa. Thus the
observables $A$ and $B$ can not be simultaneously measured to arbitrary
precision. Measuring one of the observables interferes with the measurement of
the other.

Heisenberg uncertainty principle can be applied to design a completely secure channel in 
quantum cryptographic communication protocols, that communicate random binary sequences (i.e., 
keys) with automatic eavesdropping detection. As a result, the perfect security of the Vernam 
cipher (i.e., one-time-pad) is an inexpensively implementable reality.

An example is the uncertainty of light polarization, which is extensive used in 
quantum cryptography. There are two kind photons, one is polarized in one of the two `rectilinear' directions, i.e., vertical (90 degrees)  or horizontal(0 degrees), and the another is polarized in one of two `diagonal' directions, i.e., 45 degrees or 135 degrees. 
The two directions of `rectilinear' photon or `diagonal' photon can be reliably distinguished
by a proper measurement. However, the two kind photons can not be distinguished because of 
the limitation of uncertainty principle. Rectilinear and diagonal polarization are 
complementary properties in the sense that measuring either property necessarily randomizes 
the other.

\subsection{EPR effect and Bell's Theorem}

Einstein-Podolsky-Rosen (EPR) effect play an important role in quantum information 
processing. It occurs when a spherically symmetric atom emits two photons in opposite 
directions toward two observers, Alice and Bob.The two photons are produced in an 
initial state of undefined polarization. But because of the symmetry of the initial 
state, the polarization of the photons, when measured, must have opposite values, 
provided that the measurements are of the same type. For example, if Alice and Bob 
both measure rectilinear polarization, they each equally likely to record either a 0 
(horizontal polarization) or a 1 (vertical), but if Alice obtains a 0, Bob will 
certainly obtain a 1 and vice versa. 

The unusual and important aspect of the EPR effect is that the polarization of both photons 
is determined as soon as, but not before, one of the photons is measured. This happens no 
matter how far apart the photons may be at the time. This `classical' explanation of the 
EPR effect is somewhat counterintuitive, and indeed all classic explanations of the EPR 
effect involve some implausible element, such as instantaneous action at a distance. Yet 
the mathematical formalism of quantum mechanics accounts for the EPR effect in a 
straightforward manner, and experiment have amply confirmed the existence of the phenomenon.

Of course, EPR effect may occur on various particles not only on photons. 
Einstein, Podolsky, and Rosen (EPR) in the their famous 1935 paper
[21] challenged the foundations of quantum mechanics by pointing
out a ``paradox.'' There exist spatially separated pairs of particles,
henceforth called EPR pairs, whose states are correlated in such a
way that the measurement of a chosen observable $A$ of one automatically
determines the result of the measurement of $A$ of the other. Since EPR pairs
can be pairs of particles separated at great distances, this leads to what
appears to be a paradoxical ``action at a distance.''
In 1964, Bell [22] gave a means for actually testing for
locally hidden variable theories. He proved that all
such locally hidden variable theories must satisfy the Bell inequality. Quantum mechanics
has been shown to violate the inequality.

Bell's theorem provides a method for checking eavesdropping. In following we give a brief reviews. For convenience, we explain Bell's theorem with spin-$\frac{1}{2}$ particle pairs.
Consider two measurable quantities $A$ and $B$, and label the (discrete) possible values 
of $A$ and $B$ by $\alpha_i$ and $\beta_j$, the corresponding  unit vectors are $a_i$,and $b_j$,
respectively. Then
$$E(a_i,b_j)=P_{++}(a_i,b_j)+P_{--}(a_i,b_j)-P_{-+}(a_i,b_j)-P_{+-}(a_i,b_j),$$
where $P_{\pm\pm}(a_i,b_j)$ denotes the probability that result $\pm 1$ has been obtained 
along  $a_i$ and $\pm 1$ along $b_j$.
The correlation coefficient $S$ is obtained
$$S=\sum_{i,j}E(a_i,b_j).$$
According to quantum rules
$$E(a_i,b_j)=-a_i\cdot b_j.$$
So, quantum mechanics requires
$$S=-2\sqrt{2}.$$
Intervention of eavesdropper induces
$$S=\int\rho(n_a,n_b) dn_a dn_b[\sqrt{2}n_a\cdot b_b],$$
where $n_a,n_b$ are two unit vectors (for particles a, and b, respectively), $\rho(n_a,n_b)$
is the probability of intercepting a spin component along a given direction for a particular
measurement. In this case, 
$$-\sqrt{2}\leq S\leq\sqrt{2}.$$
By the correlation coefficients, the legitimate user may detect the eavesdropper.

\section{Quantum key distribution protocols}

A lot of QKD protocols have been presented, they have similar procedure. The legitimate communicators, known as Alice and Bob, communicate over a public channel in 
following phases. 
step 1 is dedicated to raw key extraction, step 2 to error estimation, step 3 to checking eavesdropping, step 4 to reconciliation, i.e., to reconciled
key extraction, and step 5 to privacy amplification, i.e.,
extraction of final secret key.

\subsection {Quantum key distribution protocols}

{\bf Step 1. Quantum transmission over quantum communication channel}

The communicators set up a quantum channel, then they transmit quantum states (Qubits) over
the quantum channel. It is noted that the transmission model is different for different QKD protocol. Two typical transmission are BB84 protocol and EPR protocol. The former transmits 
non-commute quantum states, and the latter transmits one of each EPR pairs\\
{\bf Step 2. Extraction of raw key  over a public channel}

After Alice and Bob obtain what is call the raw data by the quantum
transmission, the raw data must be sifted because it consists of those bits
which Bob either did not receive at all or did not correctly measure in the
basis used to transmit them.  Such ``non-receptions'' could be caused by Eve's intrusion 
or by dark counts in Bob's detecting device. The location of the dark counts are, of course,
communicated by Bob to Alice over the public channel. By comparison publicly the basis 
between Alice and Bob, the data sifting procedure is completed.\\
{\bf Step 3. Check of eavesdropper}

This step depends on the different QKD protocols. In BB84 protocol, Alice and Bob now use 
the public channel to estimate the error rate in raw
key. They publicly select and agree upon a random sample of raw key, publicly
compare these bits to obtain an estimate $R$ of the error-rate. These revealed
bits are discarded from raw key. If $R$ exceeds a certain threshold $R_{Max}$,
then it will be impossible for Alice and Bob to arrive at a common secret key.
If so, Alice and Bob return to stage 1 to start over. On the other hand, If
the error estimate $R$ does not exceed $R_{Max}$, then Alice and Bob move onto
phase 3. In EPR protocol, one may use the correction of EPR pairs to check eavesdropping.\\
{\bf Step 4. Extraction of reconciled key}

In step 2, Alice and Bob's objective is to remove
all errors from what remains of raw key to produce an error free common key,
called reconciled key. This phase is of course called
reconciliation, and takes place in two stage.

In stage 1, Alice and Bob publicly agree upon a random permutation, and apply
it to what remains of their respective raw keys. Next Alice and Bob partition
the remnant raw key into blocks of length $\ell$, where the length $\ell$ is
chosen so that blocks of this length are unlikely to contain more than one
error. For each of these blocks, Alice and Bob publicly compare overall parity
checks, making sure each time to discard the last bit of the compared block.
Each time a overall parity check does not agree, Alice and Bob initiate a
binary search for the error, i.e., bisecting the block into two subblocks,
publicly comparing the parities for each of these subblocks, discarding the
right most bit of each subblock. They continue their bisective search on the
subblock for which their parities are not in agreement. This bisective search
continues until the erroneous bit is located and deleted. They then continue
to the next $\ell$-block.

Stage 1 is repeated, i.e., a random permutation is chosen, remnant raw key is
partitioned into blocks of length $\ell$, parities are compared, etc. This is
done until it becomes inefficient to continue in this fashion.

Alice and Bob then move to stage 2 by using a more refined reconciliation
procedure. They publicly select randomly chosen subsets of remnant raw key,
publicly compare parities, each time discarding an agreed upon bit from their
chosen key sample. If a parity should not agree, they employ the binary search
strategy of step 1 to locate and delete the error.

Finally, when, for some fixed number $N$ of consecutive repetitions of stage 2,
no error is found, Alice and Bob assume that to a very high probability, the
remnant raw key is without error. Alice and Bob now rename the remnant raw
key reconciled key, and move on to the final and last phase of their communication.\\
{\bf Step 5. Privacy amplification}

Alice and Bob now have a common reconciled key which they know is only
partially secret from Eve. They now begin the process of privacy
amplification, which is the extraction of a secret key from a partially
secret one [23].

By the distillation art of secret key, the so called privacy amplification,
a final secure quantum key is generated and distributed. The basic principle of
privacy amplification is as follows. Let Alice and Bob shared a random
variable $W$, such as a random $n$-bit string, while an eavesdropper Eve learns a
corrected random variable $V$, providing at most $t<n$ bits of information about
$W$, i.e., $H(W|V)\leq n-t$. Eve is allowed to specify an arbitrary distribution
$P_{VW}$ (unknown to Alice and Bob) subject to the only constraint that $R(W|
V=v)\leq n-t$ with high probability (over values $v$), where $R(W|V=v)$ denotes the
second-order conditional Renyi entropy of $W$, given $V=v$. For any $s<n-t$,
Alice and Bob can distill $r=n-t-s$ bits of the secret key $K=G(W)$ while keeping
Eve's information about $K$ exponentially small in $s$ , by publicly choosing the
compression function $G$ at random from a suitable class of maps into $\{0,1\}^
{n-t-s}$. It can be shown that Eve's average information about the final secret key
is less than $2^{-s}/ln2 $ bits.

\subsection{ The drawback of previous QKD protocols}

Obviously, the above procedure is based on the legitimate users, refereed to as
Alice and Bob. However, the practical existence of impersonation of
Alice or Bob by eavesdropper, make us have to take some action to against the
eavesdropper, an efficient way is to verify the communicators' identity. 
Unfortunately, there is no known way to initiate authentication without
initially exchanging secret key over a secure communication channel in previous protocols. 

In fact, quantum key distribution protocol is completely insecurity under the attacking of 
middle-attack. When Alice communicates Bob, Eve intercepts all qubit sent by Alice, and communicates Bob with impersonating Alice. Finally, Eve obtains two keys $K_{AE}, K_{EB}$,
where $K_{AE}$ represents the secret key between Alice and Eve, and $K_{AE}$ represents the 
secret key between Bob and Eve. As a result Eve can easily decrypt the ciphertext sent by 
Alice or Bob. 

Of course, Alice and Bob may use the classic (where `classic' contraposes quantum) 
authentication technology to prove 
the legitimated identity. However, because Alice and Bob can not simultaneously complete the identity verification and quantum key distribution, Eve may avoid the authentication procedure.
So, practically, QKD protocol with identity verification is necessary. 
In the follows, we improve the previous quantum key distribution scheme to
guarantee the security of quantum key for truly legitimate users.

\section{QKD protocol with identity verification}

Follows we propose a scheme to implement quantum authentication in QKD protocol. It may be 
implemented by non-commute quantum states or non-orthogonal quantum states with Heisenberg 
uncertainty principle. It also can be implement by EPR pairs associated with Bell's theorem.
In this paper, we use EPR pair with the Bell's theorem to implement quantum authentication. 
Both the identity verification and quantum key distribution are used in our proposed secure 
authentication protocol. There are two phase in the quantum authentication protocol. The initial 
phase is completed at the key information center to set up the system. and the authentication 
phase is executed between the two communication parties to achieve mutual authentication and 
exchange the secure quantum key.

\subsection{Initial Phase}

Assuming the information center is legitimate and believable. The information center is 
responsible neither for mutual authentication nor for the generation of quantum keys. The role of this center is to simply help the legitimate user to obtain the authentication key. In initial phase, we use the Biham's technology[24]. It uses the quantum memory, about the implementation of 
qubit in quantum memory may refer to reference[24]. In fact, the communicators and the 
center are composed of a network. 
When the secure network system is setting up, the information center will execute the following
steps. 
\begin{enumerate}
\item Alice and Bob send the center their $ID_A,ID_B$ to register to this secure network.
Then the center sets up quantum channel between Alice and the center, and between Bob and the center. 
\item Alice and Bob respectively prepare EPR pairs, and send respectively one of the each EPR 
pairs to the center. 
\item Check eavesdropping between Alice and the center, and between Bob and the center.
The center randomly chooses qubits from the strings sent by Alice and Bob, and checks the correction of quantum states like in EPR protocol.
\item The center must be able to keep the quantum states for a while ( in case the states do not arrive at the same time from Alice and Bob) and then measures the eigenvalue of the 
total-spin operator of the first pair, the second pair. etc. Except for the qubits for detection 
of eavesdropping.
\item The center tells Alice and Bob the result of the measurement. 
\item If the result of the measurement is $s=0$, Bob knows Alice's bit and vice versa. If the 
result of the measurement is $s=1$, Alice and Bob discard the transmission.
\item Alice and Bob keep the bits which correspond to $s=0$ as the raw authentication key $K'_1$. 
Proceeding the key $K'_1$ like QKD described in section 3, one obtains the authentication
key $K_1$.
\end{enumerate}

Once the legitimate users obtain the authentication keys $K_1$, the information center is not 
needed in further communication between Alice and Bob.

\subsection{Authentication phase}

Step 1. Alice and Bob transfer the common key $K_1$ into a sequence of measurement basis.
While Alice and Bob need to verify their identification, or need to set up a new communication, 
they secretly transfer the reserved 
common key into a sequence of measurement basis. For example, if Alice and Bob use the 
measurement basis of polarization photon which was used in BB84 quantum key distribution protocol, they may let the bit '0' correspond to rectilinear measurement basis and '1' 
correspond to diagonal measurement basis, or vice versa.
We represent rectilinear measurement basis by the symbol $\oslash$, and represent diagonal measurement basis by the symbol $\odot$. After transferred, Alice and Bob obtain a sequence of measurement basis, respectively. For example, if  the common key is 001101, the sequence of measurement basic is $\odot\odot\oslash\oslash\odot\oslash$.

Step 2. Alice and Bob set up a quantum communication channel. When Alice wants to secretly 
communicate Bob, Alice and Bob need to set up a quantum channel. The transmitting quantum 
states in the quantum channel may be arbitrary. For example the polarization photon state 
or the phase correction states. In this protocol, we use the phase correction states. So the 
channel consists of a source that emits EPR pairs of spin-$\frac{1}{2}$ particle, in a singlet 
state. The particles fly apart along the $z$ axis, towards the two legitimate users of the channel. Alice chooses a random  basis for measuring one numbering of each EPR pair 
of particles. The other particle of each EPR pair is measured by Bob in the next step.
Alice's measurement results in effect determine, through  the EPR corrections, a sequence of states for Bob's particles.

Step 3. Bob measures the stings of quantum states.
Bob randomly measures the sequence of quantum states by using two 
measurement basis $M, M_{K_1}$, where $M$ is the measurement basis for quantum key 
distribution and for obtaining new authentication key, $M_{K_1}$ is the measurement basis 
for identity authentication in the current communication. $M$ is like that in EPR protocol.

Step 4. Alice and Bob check the eavesdropper. 
For secure communication, the legitimates users Alice and Bob need to firstly detect 
eavesdropper. Bob randomly chooses some measurement results measured by the basis $M$ for 
checking the correction of EPR pair. According to the Bell's theorem described in subsection 
2.2 to judge the eavesdropping. 

Step 5. Bob encrypts his results measured by $M_{K_1}$.
Although Bob does not know the qubits measured by Alice, it will not influence the identity verification. Expressing the strings of quantum states for authentication by 
$$|\Psi>=\{|\psi_1>, |\psi_2>,\cdots, |\psi_n>\}.$$
Where $|\psi_i>$ represents a qubit received by Bob. After measurement, Bob obtains
$$|\Phi>=M_{K_1}|\Psi>,$$ 
where $|\Phi>=\{|\phi_1>, |\phi_2>,\cdots, |\phi_n>\}$ represents the measurement results 
under measurement basis $M_{K_1}$. Transferring $|\Phi>$ into binary bits strings $m$, and then using $K_1$ to encrypt it, Bob obtains the ciphertext 
$$y=E_{K_1}(m).$$ 
Bob sends Alice the ciphertext $y$, and tells Alice the corresponding sequence numbers of
quantum states $|\psi_i>, i=1,2,\cdots, n$. 

Step 5. Verifying Bob's identity.
Having received Bob's results, Alice analyzes Bob's results. Alice decrypts the ciphertext, $$m=E^{-1}_{K_1}(y),$$ 
and compare her results with $m$, thereby Alice gets the measurement basis $M_{K_t}$. If
$K_t=K_1$, Bob's identity is true. 

Step 6 Verifying Alice' identity.
After Alice decrypted the ciphertext, Alice sends Bob the result $m'$. If $m'=m$, the 
Alice's identity is true. 

Step 7. Alice and Bob distribute the quantum secret key. If the communicators are 
legitimate, Alice and Bob distribute the quantum secret key using the remainder Qubits. 
The process is same as EPR protocol.

Step 8. Alice and Bob discard the authentication keys $K_1$, and set up new 
authentication keys. After authentication, the authentication key $K_1$ is no longer use. The 
legitimate users obtain a new authentication key. The method is like that used in QKD protocol. 
Of course, one can directly take portion bits from the final quantum key as the authentication 
key.

It has been noted that the presented protocol can not prevent voluntary attack. This is a 
drawback of quantum cryptography. How to prevent the voluntary attack needs further 
investigation.

\section{Security analysis}

The proposed QKD protocol with authentication is provably secure.
Obviously, the QKD is provably secure because we use the previous QKD protocol. So in 
following, we mainly analyze the security of authentication scheme. 
In initial phase, the security derives from the security of the EPR protocol, and relies on 
the fact that the singlet state is the only state for which the two spins are anticorrelated 
both in $\hat{S}_z$ and in the $\hat{S}_x$ basis. So eavesdropper and the center can not 
obtain the authentication key $K_1$.

In authentication phase, we believe this scheme is secure as the follows reason.
i) Our protocol does not have the conspiracy problem of masquerading. If a forger wants to 
masquerade user Alice or Bob to communicate with others, he must find the common key. However,
it is difficult to obtain the shared common secret because of the follows two reasons. 
First, the authentication key is obtained by the quantum key distribution protocol which is provably secure, so the authentication key is secure. Second the authentication key is used 
only one times, eavesdropper does not know any information about the authentication key.  
ii) The replay-attack will also not succeed in our protocol because the key is used only one 
times. iii) The quantum attacking strategy is invalid, the reason is the same as the analysis 
for previous QKD protocols.

There is a weakness in our protocol. The only weakness of our 
protocol is the reservation of authentication key. Although the obtaining of the common
key in the last quantum communication is provably secure, the common key reservation has not circumvented the catch 22 problem. In fact, this drawback exists in all symmetric 
cryptographic authentication system.  Of course, we can use the EPR effects or other quantum effect to keep the common key, but the reservation time is very short according to current 
technology. A long time correction of quantum states is need in the future.

\section{Conclusion}

The previous QKD protocols are based on the legitimate users. However, the practical 
existence of impersonation of Alice or Bob by eavesdropper, make us have to take some 
action to against the eavesdropper, an efficient way is to verify the communicators' identity. 
Unfortunately, there is no known way to initiate authentication without
initially exchanging secret key over a secure communication channel in previous protocols. Of 
course, one can use the classic authentication protocol to verify identity. However, because 
the authentication and the QKD can not be simultaneous, Eve can escape the authentication 
procedure. In addition, quantum key distribution protocol is completely insecurity under the attacking of middle-attack. 

In this paper, we proposed a QKD protocol that can simultaneously distribute the quantum 
secret key and verify the communicators' identity. The QKD is implemented by the 
previous EPR protocol, the authentication is implemented by the symmetric cryptographic scheme
with quantum effects. The presented scheme is provably secure.

we use EPR effects with Bell' theorem to implement quantum authentication. 
It can prevent impersonation and middle-attack. Of course, it can also be 
implemented by noncommute quantum states or non-orthogonal quantum states with 
Heisenberg uncertainty principle. 

\begin{flushleft}
{\bf Acknowledgments}
\end{flushleft}

This project was supported by the National Natural Science Foundation of China, Grant no: 69803008.

\eject
\begin{enumerate}
\begin{flushleft}
{\bf References}\\[0.1cm]
\end{flushleft}
\item S.Wiesner,  Conjugate coding, Sigact News, vol. 15, no. 1, 1983, pp. 78 - 88; original manuscript written circa 1970. 
\item C. H.Bennett, G.Brassard, S.Breidbart, and S.Wiesner,  Quantum cryptography, or 
unforgeable subway tokens, Advances in Cryptology: 
Proceedings of Crypto 82, August 1982, Plenum Press, New York, pp. 267 - 275. 
\item C. H. Bennett, and G.Brassard,  An update on quantum cryptography, Advances in 
Cryptology: Proceedings of Crypto 84, August 1984, Springer - Verlag, pp. 475 - 480. 
\item C. H.Bennett,  Quantum cryptography using any two nonorthogonal states, Physical 
Review Letters, vol. 68, no. 21, 25 May 1992, pp. 3121 - 2124. 
\item A. K.Ekert, Quantum cryptography based on Bell's theorem, Physical Review Letters,
 vol. 67, no. 6, 5 August 1991, pp. 661 - 663. 
A.K.Ekert, J.G.Rarity, P.R.Tapster, and G.M.Palma, Practical quantum cryptography based on two-photon interferometry, Phys. Rev. Lett. 69, 1293(1992). 
\item C.H.Bennett,F.Bessette, G.Brassard, L.Salvail and J.Smolin, Experimental quantum 
cryptography, J.Cryptology 5, 3 (1992). 

\item J.Breguet, A.Muller, and N.Gisin, Quantum cryptography with polarized photons in optical fibres,  Journal of Modern Optics, 41, no.12, 2405-2412(1994).

\item S.Phoenix, S.Barnett, P.Townsend and K.Blow, Multi-user quantum cryptography on optical networks, Journal of Modern Optics, 42, no.6, 1155-1163(1995).
\item S. M.Barnett,  and S. J. D.Phoenix,  "Bell's inequality and rejected-data protocols 
for quantum cryptography", Journal of Modern Optics, vol. 40, no. 8, August 1993, 
pp. 1443 - 1448. 
\item C. H.Bennett,  and G.Brassard,  "Quantum cryptography and its application to provably secure  key expansion, public-key distribution, and coin-tossing", IEEE International 
Symposium on Information Theory, September 1983, page 91. 

\item B.Hutter, A.Ekert, Information gian in quantum eavesdropping, Journal of Modern 
Optics, 41, 2455-2466(1994). 
\item C.A.Fuchs, N.Gisin, R.B.Griffiths, C.S.Niu, and A.Peres, Optimal eavesdropping in quantum cryptography, Phys. Rev. A 56, 1163-1172 (1997).
\item B.A.Slutsky, R.Rao, P.C.Sun, and Y.Fainman, Security of quantum cryptography against individual attacks, Phys. Rev. A 57, 2383(1998).
\item Brandt, Howard E., John M. Meyers, And Samuel J.
Lomonaco,Jr., Aspects of entangled translucent eavesdropping in
quantum cryptography, Phys. Rev. A, Vol. 56, No. 6, December 1997, pp. 4456 - 4465
\item C.Niu, and R.Griffiths, Optimal copyinf of one quantum bit, Phys.Rev.A, Vol.58, no. 6,
4377-4393,(1998).
\item L.Duan and G.Guo, Probabilistic cloning and identification of Linearly independent 
quantum states, Phys. Rev. Lett., Vol.80 no.22, 4999-5002,(1998).

\item J.G.Rarity, P.C.M.Owens, and P.R.Tapster, Quantum random-number generation and key 
sharing, Journal of Modern Optics, 41, 2435(1994). 
\item P.D.Townsend, J.G.Rarity, and P.R.Tapster,  Single photon interference in a 10 
km long optical fibre interferometer, Electronics Letters, vol. 29, no. 7, April 1993, pp. 
634 - 635;
\item P.D.Townsend, J.G.Rarity, and P.R.Tapster, Enhanced single photon fringe 
visibility in a 10 km-long prototype quantum cryptography channel, Electronics Letters, 
vol. 29, no. 14, 8 July 1993, pp. 1291 - 1293. 
\item C.Marand and P.D.Townsend, Quantum key distribution over distances as long as 30 Km, 
Optics Letters, Vol. 20, no. 16, 1695-1697(1995).
\item A.Einstein, B.Podolsky and N.Rosen, Can quantum mechanical description of physical reality be considered complete? Phys. Rev. 47, 777(1935)
\item J.S.Bell, Physics (Long Island City, N.Y.) 1, 195(1965).
\item C.H.Bennett, G.Brassard, C.Crepeau and U.M.Maurer, Generalized privacy amplification, IEEE Trans. Inform. Theory, 41, 1915(1995).
\item E.Biham, B.Huttner, and T.Mor, Quantum cryptographic network based on quantum memories,
Phys.Rev.A, Vol. 54, no. 4, 2651-2658, (1996).

\end{enumerate}
\end{document}